\documentclass[a4paper]{article}

\usepackage{INTERSPEECH2022}
\usepackage{cite,bm}
\usepackage{amssymb}
\usepackage{amsfonts}
\usepackage{comment}
\usepackage{amsmath,mathtools}
\usepackage{algorithm}
\usepackage[noend]{algpseudocode}
\usepackage{booktabs}
\usepackage{flushend}
\usepackage{tabularx}
\usepackage[binary-units]{siunitx}
\usepackage{threeparttable}
\pdfoutput=1

\AtBeginDocument{
  \abovedisplayskip     =0.5\abovedisplayskip
  \abovedisplayshortskip=0.5\abovedisplayshortskip
  \belowdisplayskip     =0.5\belowdisplayskip
  \belowdisplayshortskip=0.5\belowdisplayshortskip
}

\title{Independence-based Joint Dereverberation and Separation \\with Neural Source Model}
\name{
    Kohei Saijo$^{1,2*}$, 
    Robin Scheibler$^2$
    \thanks{
        *This work was done during an internship at LINE Corporation.
    }
}
\address{
  $^1$Waseda University, Japan \;  $^2$LINE Corporation, Japan}
\email{saijo@pcl.cs.waseda.ac.jp}

\begin{document}
%\ninept
%
\maketitle
\begin{abstract}
We propose an independence-based joint dereverberation and separation method with a neural source model.
We introduce a neural network in the framework of time-decorrelation iterative source steering, which is an extension of independent vector analysis to joint dereverberation and separation.
The network is trained in an end-to-end manner with a permutation invariant loss on the time-domain separation output signals.
Our proposed method can be applied in any situation with at least as many microphones as sources, regardless of their number.
In experiments, we demonstrate that our method results in high performance in terms of both speech quality metrics and word error rate (WER), even for mixtures with a different number of speakers than training.
Furthermore, the model, trained on synthetic mixtures, without any modifications, greatly reduces the WER on the recorded dataset LibriCSS. % by XX\% compared to a model Y trained with double the data and 4 times the number of parameters.
\end{abstract}

\section{Introduction}
\label{sec:intro}
Since speech recordings are generally contaminated by interference, background noise, and reverberation, source separation and dereverberation have been studied as pre-processing for speech systems, e.g., automatic speech recognition (ASR).
On the one hand, blind source separation (BSS) such as independent component analysis (ICA)~\cite{ica}, independent vector analysis (IVA)~\cite{iva_kim,iva_hiroe}, independent low-rank matrix analysis (ILRMA)~\cite{ilrma}, dereverberation techniques such as weighted prediction error (WPE)~\cite{wpe}, and their joint optimization such as ILRMA-T~\cite{ilrma-t} have been an area of intense research.
On the other hand, supervised learning of deep neural networks (DNN) has been proposed for the single-channel source separation~\cite{dc,pit,tasnet}.
It has also been extended to the multi-channel setup, where the time-frequency (TF) masks or separated signals estimated by the single-channel network are used to obtain spatial statistics for beamforming~\cite{mvdr_heymann,mvdr_ochiai,cisdr}.
Such linear filtering techniques have empirically been shown to be better as a front-end to ASR than non-linear separation approaches~\cite{haeb2020far}.

We focus on linear dereverberation and separation techniques.
Among them, the cascade connection of dereverberation and separation such as WPE~\cite{wpe} and neural beamforming~\cite{mvdr_heymann,mvdr_ochiai,cisdr} or their joint otimization~\cite{cbf} have shown good performance in both speech metrics and WER.
However, most of them have two drawbacks.
First, they are sensitive to domain mismatch, because they highly depend on the front-end mask estimation network.
Second, the number of sources to be separated is fixed.
Most methods prepare an output layer for the number of sources, which is practically inconvenient because the number of sources is usually unknown in advance.

BSS with a neural source model~\cite{duong_dnn,idlma,auxiva-iss-dnn} is one of the methods that work regardless of the number of sources.
In~\cite{duong_dnn,idlma}, a pre-trained network is utilized as the source model.
End-to-end training of the source model of IVA was also proposed in~\cite{auxiva-iss-dnn}, where stable training was enabled thanks to the inverse-free spatial model update for IVA, i.e., iterative source steering (ISS)~\cite{iss}.
However, \cite{auxiva-iss-dnn} tackled only separation, not dereverberation.
%In addition, training iterative methods with backpropagation consumes memory linearly in the number of iterations, which makes training difficult with limited resources.
Recently, an extension of ISS to joint dereverberation and separation, time-decorrelation ISS (T-ISS)~\cite{ilrma-t-iss} has been proposed.
Still, in \cite{ilrma-t-iss}, the conventional non-negative low-rank model~\cite{ilrma, ilrma-t}, which has no trainable parameters, was used as source model, limiting the performance.

%---
In this work, we propose to replace the non-trainable source model in the original T-ISS with a trainable neural network.
We train this neural source model in an end-to-end manner with a permutation invariant loss~\cite{cisdr} on the time-domain separation output signals.
Although end-to-end training of iterative methods requires memory proportional to the number of iterations, we introduce a memory-efficient gradient computation technique, \textit{demixing matrix checkpointing} (DMC), a derivative of checkpointing technique~\cite{checkpointing} which makes the memory cost nearly constant regardless of the number of iterations.
Our proposed method is robust to domain mismatch because the trained model can be applied whenever there at least as many microphones as sources, regardless of their number.
While the neural source model extracts a single source from a single channel input, the parameter-free spatial model update of T-ISS handles the difference in the number of microphones or sources.

The key contributions are summarized as follows. 
1)~This is the first work to train a neural source model for T-ISS.
We show the superiority of our proposed method to T-ISS with conventional models~\cite{ilrma-t-iss} and cascade connection of dereverberation~\cite{wpe} and separation~\cite{cisdr, auxiva-iss-dnn}, regardless of the number of speakers.
2)~We demonstrate that the proposed method trained with fully overlapped synthetic mixtures also works well on the low-overlap recordings of LibriCSS~\cite{libricss} without any modification.
Our proposed method outperformed a strong Conformer-based MVDR beamforming approach~\cite{libricss_conformer}.

\section{Background}
\label{sec:background}

Assuming $N$ sources are captured by $M$ microphones, the microphone input signal in the short-time Fourier transform (STFT) domain is modeled auto-regressively, namely,
%------------------------------
\begin{align}
  \label{eqn:signal_model}
    \bm{x}_{f,t} = {\bm{A}_f}{\bm{s}_{f,t}} + \sum_{l=D+1}^{D+L} {\bm{Z}_{f,l} \bm{x}_{f,t-l}},
\end{align}
%------------------------------
where $\bm{s}_{f,t}$ is the clean sources vector, $\bm{A}_{f}\in\mathbb{C}^{{M}\times{N}}$ the mixing matrix, and ${\bm{Z}_{f,l}}$ produces the reverberation component from past samples. 
$L$ is the tap length and $D$ is the delay to separate the direct signal and reverberation.
${f} = {1,\ldots,F}$ and ${t} = {1,\ldots,T}$ are the frequency bin and the time frame index in STFT-domain. 
We can rewrite \eqref{eqn:signal_model} as
%------------------------------
\begin{align}
  \label{eqn:signal_model_simple}
    \bm{x}_{f,t} = {\bm{A}_f}{\bm{s}_{f,t}} + \bm{\bar{Z}}_{f}\bm{\bar{x}}_{f,t},
\end{align}
%------------------------------
where $\bm{\bar{Z}}_{f} = [\bm{Z}_{f,D+1},\dots,\bm{Z}_{f,D+L}]\in\mathbb{C}^{{M}\times{ML}}$ and  $\bm{\bar{x}}_{f,t} = [\bm{{x}}_{f,t-D-1}^\top,\dots,\bm{{x}}_{f,t-D-L}^\top]^\top{}\in\mathbb{C}^{ML}$.
Supposing we know $\bm{W}_f=\bm{A}_f^{-1}$, i.e., the demixing matrix, and $\bm{\bar{Z}}_f$, we recover
%------------------------------
\begin{align}
  \label{eqn:separation}
    \bm{s}_{f,t}=\bm{W}_{f}(\bm{x}_{f,t}-\bm{\bar{Z}}_{f}\bm{\bar{x}}_{f,t}).
\end{align}
%------------------------------
Here we assume the determined case, i.e., $M=N$. 
In the following, $\bm{I}$ and $\bm{e}_{n}$ denote the identity matrix and the $n$th canonical basis vector, $^{*}$ the complex conjugate, and $^\top{}$ and $^{\mathsf{H}}$ the transpose and Hermitian transpose of vectors or matrices.

Since conventional dereverberation methods such as WPE~\cite{wpe} assume the existence of only one source, the cascade connection of dereverberation and separation is not optimal.
To circumvent this problem, a \textit{joint} dereverberation and separation framework was introduced as part of ILRMA-T~\cite{ilrma-t}, where the estimated clean sources vector $\bm{y}_{f,t}$ is obtained with a unified filter $\bm{P}_{f}=\bm{W}_{f}[\bm{I}, -\bm{\bar{Z}}_{f}]\in\mathbb{C}^{{M}\times{M(L+1)}}$
as $\bm{y}_{f,t}=\bm{P}_{f}{\bm{\tilde{x}}_{f,t}}$ where ${\bm{\tilde{x}}_{f,t}} = [\bm{x}_{f,t}^\top{}, \bm{\bar{x}}_{f,t}^\top]^\top{}\in\mathbb{C}^{M(L+1)}$. 
Since both $\bm{y}_{f,t}$ and $\bm{P}_{f}$ are unknown, we solve it by maximum likelihood with a generative model of the sources.
ILRMA-T assumes that each TF bin of the $n$th source, $y_{n,f,t}$, belongs to a complex Gaussian distribution, $p(y_{n,f,t}) = \frac{1}{\pi r_{n,f,t}} \exp\left(-\frac{|y_{n,f,t}|^2}{r_{n,f,t}}\right)$, where $r_{n,f,t}$ is the time-varying variance of the $n$th source and modeled as low-rank non-negative~\cite{ilrma}.
Assuming each source is independent, we obtain the following negative log-likelihood~\cite{ilrma-t},
%The unified filter can be estimated by minimizing the following negative log-likelihood cost function~\cite{ilrma-t},
%------------------------------
\begin{align}
  \label{eqn:ilrma-t-iss-cost}
    \mathcal{J} = \sum_{f,t} \left[-2 \log|\mathrm{det}{\bm{W}_f}| + 
    %\frac{1}{T} \sum_{n} {u_{f,t}(\bm{Y}_{n}) |\bm{p}^{\mathsf{H}}_{n,f}\bm{\tilde{x}}_{f,t}|^2}\right],
    \frac{1}{T} \sum_{n} \frac{|\bm{p}^{\mathsf{H}}_{n,f}\bm{\tilde{x}}_{f,t}|^2}{r_{n,f,t}}\right],
\end{align}
%------------------------------
where $\bm{p}^{\mathsf{H}}_{n,f}$ is the $n$th row vector of $\bm{P}_{f}$.
For the sake of generality, we hereafter refer to the reciprocal of the time-varying variance as the function of $\bm{y}$, i.e., $u_{f,t}(\bm{Y}_{n}) = 1 / r_{n,f,t}$, where $(\bm{Y}_{n})_{f,t}=(\bm{y}_{f,t})_{n}$.
The unified filter $\bm{P}_{f}$ can be estimated by iteratively minimizing (\ref{eqn:ilrma-t-iss-cost}).
To avoid large matrix inversions in the update rule derived from iterative projection algorithm~\cite{auxiva_ip,ilrma-t}, the inverse-free rank-1 update rule, T-ISS has been proposed~\cite{iss,ilrma-t-iss}.
For $1 \leq n \leq M$, the update is done as,
%------------------------------
\begin{align}
  \label{eqn:ilrma-t-iss-rank1update1}
    \bm{P}_{f} \gets \bm{P}_{f}-\bm{v}_{n,f}\bm{p}^{\mathsf{H}}_{n,f},
\end{align}
%------------------------------
%where $\bm{v}_{n,f}=[v_{1,n,f},\dots,v_{N,n,f}]^\top{}$ is chosen to minimize \eqref{eqn:ilrma-t-iss-cost} and $\bm{p}^{\mathsf{H}}_{n,f}$ is the $n$-th row vector of $\bm{P}_{f}$, which yields,
where $\bm{v}_{n,f}=[v_{1,n,f},\dots,v_{M,n,f}]^\top{}$ is chosen to minimize \eqref{eqn:ilrma-t-iss-cost}, which yields,
%------------------------------
\begin{align}
    \label{eqn:ilrma-t-iss-update1}
    v_{m,n,f} = 
    \begin{cases}
    1 - \left(\frac{1}{T} \sum_{t} u_{f,t}(\bm{Y}_n) |y_{n,f,t}|^2 \right)^{-\frac{1}{2}}, & \text{if $m=n$,} \smallskip \\
    \frac{\sum_{t} u_{f,t}(\bm{Y}_m) y_{m,f,t}y^{*}_{n,f,t}}{\sum_{t} u_{f,t}(\bm{Y}_m) |y_{n,f,t}|^2 }, & \text{otherwise}.
    \end{cases}
\end{align}
%------------------------------
For $n>M$, the update is $\bm{P}_{f} \gets \bm{P}_{f}-\bm{v}_{n,f}\bm{e}_{n,f}^\top{}$ and $v_{m,n,f}$ is,
%------------------------------
\begin{align}
    \label{eqn:ilrma-t-iss-update2}
    v_{m,n,f} = 
    \frac{\sum_{t} u_{f,t}(\bm{Y}_m) y_{m,f,t}\tilde{x}^{*}_{n,f,t}}
    {\sum_{t} u_{f,t}(\bm{Y}_m) |\tilde{x}_{n,f,t}|^2 },
\end{align}
%------------------------------
where $\tilde{x}_{n,f,t}$ is the $n$th component of $\bm{\tilde{x}}_{f,t}$.

\section{Proposed source model learning}
\label{sec:proposed_method}

%------------------------------
\begin{figure}[t]
\centering
\centerline{\includegraphics[width=\linewidth]{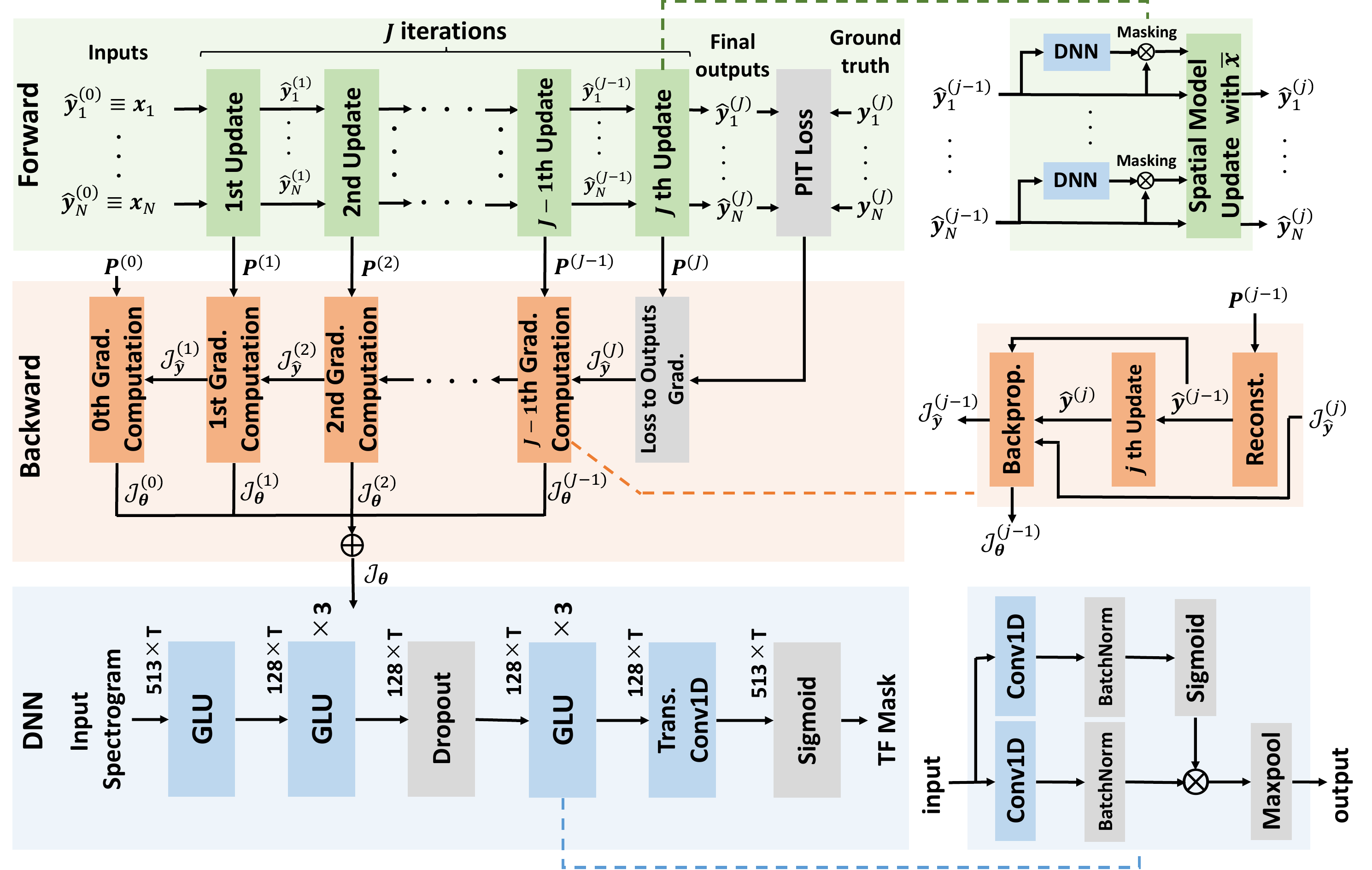}}
%\vspace{-1mm}
\caption{Overview of our proposed method. We train a source model DNN (blue) directly through $J$ iterations with a memory-efficient gradient computation. Unified filter $\bm{P}^{(j)}$ is stored in the forward pass (green) and used to reconstruct the separated signals in the backward pass (orange).
The gradient to update the DNN parameter $\mathcal{J}_{\bm{\theta}}$ is simply obtained by summing up the gradient at each iteration. 
}
\label{fig:overview}
%\vspace{-1mm}
\end{figure}
%------------------------------

Conventionally, the source mask $u_{f,t}(\bm{Y}_n)$ is derived from a probabilistic model of speech such as low-rank non-negative model~\cite{ilrma-t}.
Instead, we propose to replace it with a neural network and learn its weights by gradient descent in the independence-based joint dereverberation and separation framework of T-ISS~\cite{ilrma-t-iss}.
%Specifically, a TF mask is estimated in each iteration by the DNN and used as $u_{f,t}(\bm{Y}_n)$.
The loss is computed between the ground truth and the final separation output of multiple iterations.
To reduce memory cost to train through multiple iterations, we introduce a memory-efficient gradient computation technique, DMC, as an alternative to backpropagation (BP).
Since the neural source model extracts a single source from a single channel input and the spatial model update of T-ISS has no trainable parameters, our proposed method can be applied to mixtures with any number of microphones or sources within an (over-) determined case.
For the same reason, the proposed method is robust to domain mismatch.
The overview of our proposed method is shown in Fig.~\ref{fig:overview}.

\subsection{Network Architecture}
\label{sec:architecture}
The source model network architecture is shown at the bottom of Fig.~\ref{fig:overview}.
It consists of multiple gated linear units (GLU)~\cite{glu} and a transpose convolution layer, all with kernel size three.
It has 2.18M trainable parameters in total.
The input is down-sampled in the first block and then fed into six GLU blocks with a dropout layer with a probability of 0.5 between the third and the fourth.
Finally, the transpose convolution layer up-samples the intermediate feature to the same size as the original input, and a TF mask is output thorough the sigmoid activation.
The mask may be understood as hiding the target source in the current source spectrogram estimate.

\subsection{Loss Function}
\label{sec:loss}
%We use CI-SDR loss~\cite{vincentPerformanceMeasurementBlind2006, cisdr} to train the source model DNN.
To train the neural source model, we use convolutive transfer function invariant~\cite{vincentPerformanceMeasurementBlind2006} signal-to-distortion ratio (CI-SDR) loss, which was shown to be effective for multi-channel enhancement~\cite{cisdr}. 
In CI-SDR, a $K$ tap filter $\bm{\alpha}\in\mathbb{R}^{K}$, where $K$ is usually set to 512, is used to make the SDR short-impulse-response-invariant.
Let the time-domain estimated signal and the ground truth be $\bm{\hat{s}}$ and $\bm{s}\in\mathbb{R}^{I}$, respectively.
Further define a matrix containing $K$ shifts of $\bm{s}$ in its columns, i.e., $\bm{\bar{S}}=[\bm{\bar{s}}_{1},\dots,\bm{\bar{s}}_{I}]^\top{}\in\mathbb{R}^{{I}\times{K}}$, with $i$th row $\bm{\bar{s}}_{i}=[s_{i-1},\dots,s_{i-K}]^\top{} \in\mathbb{R}^{K}$, where $i=1,\dots,I$ is the time index.
Then, let $\bm{\alpha}=(\bm{\bar{S}^\top{}}\bm{\bar{S}})^{-1}\bm{\bar{S}^\top{}}\bm{\hat{s}}$, the CI-SDR loss is,
%------------------------------------------
\begin{align}
  \label{eqn:cisdr}
    \mathcal{L}_{\rm{CI-SDR}} = -10\log_{10}{\left(\frac{||\bm{\bar{S}}\bm{\alpha}||^2}{||\bm{\bar{S}}\bm{\alpha}-\hat{\bm{s}}||^2}\right)}.
\end{align}
%------------------------------------------

\subsection{Demixing Matrix Checkpointing}
\label{sec:checkpointing}

%BP needs all the intermediate results to compute the gradient.
BP starts from the final loss and works its way backward through the computational graph computing the gradient along the way.
Thus, BP needs all the intermediate results and memory cost becomes a bottleneck for iterative framework. 
However, by storing the demixing matrices during the forward pass, we can recreate the intermediate separation results.
Based on this idea, we introduce a memory-efficient gradient computation technique, demixing matrix checkpointing (DMC). 
The backward pass of DMC is shown in Fig.\ref{fig:overview}.
%Pseudo-code for DMC is shown in Algorithm~\ref{alg:checkpointing}. 
DMC proceeds by computing the contribution of each iteration to the gradient in reverse order.
At $j$th iteration, we first reconstruct the separated signals of the previous iteration $\bm{\hat{y}}^{(j-1)}$ using $\bm{P}^{(j-1)}$ stored during the forward pass.
Then, we re-compute $\bm{\hat{y}}^{(j)}$ in the same way as the forward pass.
The gradient of the model parameter $\bm{\theta}$ at $j$th iteration, $\mathcal{J}^{(j-1)}_{\bm{\theta}}$, is computed by BP.
Since the DNN parameters are shared in all the iterations, the gradient to update the DNN parameters is obtained by summing up $\mathcal{J}^{(j)}_{\bm{\theta}}$ for all $j$.

\section{Experiments}
\label{sec:experiments}

Experimental comparisons were conducted to verify the effectiveness of our proposed method.
The baselines were as follows.
\textbf{WPE + DNN-MVDR}~\cite{wpe, cisdr}: 
Cascade connection of WPE and neural MVDR beamforming. Here, we used the BLSTM-based DNN from~\cite{cisdr} rather than the network shown in Fig.~\ref{fig:overview}.
\textbf{WPE + DNN-IVA}~\cite{wpe, auxiva-iss-dnn}: 
Cascade connection of WPE and IVA with a neural source model. 
\textbf{ILRMA-T-ISS}~\cite{ilrma-t-iss}: 
T-ISS with low-rank non-negative source model~\cite{ilrma}. 
It had no trainable parameters and performs joint dereverberation and separation blindly. 

All the algorithms were implemented in Pytorch~\cite{pytorch}.
Allocated memory and computational time were measured with the \textsf{nvidia-smi} command and the profiler of Pytorch.
Experiments were conducted on a Linux workstation with an Intel\textregistered\ Xeon\textregistered\ Gold 6230 CPU \@ \SI{2.10}{\giga\hertz} with 8 cores, an NVIDIA\textregistered\ Tesla\textregistered\ V100 graphical processing unit (GPU), and \SI{64}{\giga\byte} of RAM.

\subsection{Datasets and Experimental Setup}
\label{sec:dataset}

\textbf{WSJ1-mix}:
We used the synthetic reverberant noisy mixtures introduced in~\cite{auxiva-iss-dnn}.
It consisted of speech from the WSJ1 corpus~\cite{wsj1} and noise from the CHIME3 dataset~\cite{chime3} sampled in 16~kHz.
The reverberation times were chosen from \SIrange{200}{600}{\milli\second}.
%Compared to~\cite{auxiva-iss-dnn}, the only difference is that we removed the DC-offset by subtracting from itself the mean value of each speech sample.
The relative power of sources was chosen from \SIrange{-5}{5}{\decibel}.
The noise was scaled to attain an SNR between \SIrange{10}{30}{\decibel}.
Training, validation, and test sets contained 37416, 503, and 333 mixtures, approximately 98.5, 1.33 and 0.85 hours of mixtures, respectively.
We created two, three, and four channels mixtures with two, three and four speakers, respectively.
Only the two-speaker mixtures were used for training, and two, three and four-speaker mixtures were used for validation and test.
To evaluate word error rate (WER), we trained an ASR system with clean anechoic signals using the \textsf{wsj/asr1} recipe from the ESPNet framework~\cite{espnet}.
WER for the clean, anechoic test set was \SI{9.25}{\percent}.
During training, the batch size was eight and the input signal length was seven seconds. 
The network parameters were optimized using the Adam optimizer~\cite{adam} with the learning rate of $1.0 \times 10^{-4}$.
For STFT, we used the Hann window with the size of 1024, and hop size was 256.
The number of iterations of T-ISS and IVA were set to 20 in training and 50, 75 and 100 for two, three and four-speaker mixtures in test.
The number of iterations for WPE was 3 both in training and test.
The delay was set to 1 in T-ISS and 3 in WPE because such setting gave the best performance.
Both used $L=5$ taps.
%To solve the scale and phase ambiguity, 
We applied projection back \cite{projback} to the final output.

\noindent
\textbf{LibriCSS}:
We used LibriCSS dataset~\cite{libricss} to evaluate the performance of the models trained with WSJ1-mix on out-of-domain real-world recorded speeches.
Utterances taken from LibriSpeech test clean set were played back from loudspeakers placed in a room, and recorded by a seven channel circular microphones with radius of 4.25~cm.
The distances from the loudspeakers and the microphones ranged from 33~cm to 409~cm.
Separation performance were evaluated by WERs of separated signals using the ASR systems provided in~\cite{libricss}.
LibriCSS contained data whose average overlap ratio ranges from 0 to 40\%, 0S, 0L, 10, 20, 30 and 40\% overlaps, where 0S/0L were overlap-free recordings with short/long inter-utterance silence.
We conducted \textit{utterance-wise} evaluation using two, three and seven channels of microphones with 50, 30 and 15 iterations, respectively.
Because basically one or two speech sources were in an observation, using three or more channels corresponded to the over-determined condition. 
When over-determined case, two separated signals with the highest power were evaluated.

\subsection{Results on WSJ1-mix dataset}
\label{sec:results_wsj}

%------------------------------ 
\begin{table}[t]
%\vspace{-1mm}
\begin{center}
\caption{Average SDR in decibels, STOI, PESQ, SRMR and WER of separated signals from WSJ1-mix test set. The number of sources equals that of channels. Training was done using only 2ch. data. The proposed method is indicated by a star ($\star$).}
\label{table:result}
\vspace{-3.5mm}
%\footnotesize
\scriptsize
% @{\extracolsep{\fill}}
\begin{tabular}{@{}llc@{~~~}c@{~~~}c@{~~~}c@{~~~}c@{}}
%\begin{tabular}{llrrrrr}
\toprule
{Ch.} & {Algo.} & {SDR$^\uparrow$} & {STOI$^\uparrow$} & {PESQ$^\uparrow$} &{SRMR$^\uparrow$} & {WER$^\downarrow$} \\

\midrule
2     & Unprocessed  &-0.4  &0.728  &1.21  &5.13  &111.4\% \\
      & WPE+DNN-MVDR &10.0  &0.875  &1.63  &6.66  &54.2\%  \\
      & WPE+DNN-IVA  &~9.8  &0.880  &1.60  &6.57  &53.2\%  \\
      & ILRMA-T-ISS  &~7.9  &0.860  &1.55  &6.45  &49.3\%  \\
      & DNN-T-ISS$^\star$  &\bf{12.3} &\bf{0.907} &\bf{1.78} &\bf{6.87} &{\bf{29.2}}\%  \\
      
\midrule
3     & Unprocessed   &-3.6 &0.640 &1.14 &4.58 &145.9\% \\
      & WPE+DNN-IVA   &~7.9  &0.850 &1.47 &6.41 &72.3\%  \\
      & ILRMA-T-ISS   &~4.4  &0.799 &1.37 &6.16 &77.9\%  \\
      & DNN-T-ISS$^\star$  &\bf{~9.7} &\bf{0.879} &\bf{1.62} &\bf{6.65} &{\bf{43.8}}\%  \\

\midrule
4     & Unprocessed  &-5.4 &0.584 &1.12 &4.12 &162.8\%\\
      & WPE+DNN-IVA  &~6.6  &0.828 &1.40 &6.25 &87.0\% \\
      & ILRMA-T-ISS  &~2.0  &0.755 &1.28 &5.81 &98.1\% \\
      & DNN-T-ISS$^\star$  &\bf{~7.7} &\bf{0.849} &\bf{1.50} &\bf{6.40} &{\bf{57.3}}\%  \\
      
\bottomrule

\end{tabular}
\end{center}
\vspace{-6mm}
\end{table}
%%------------------------------

%------------------------------
\begin{figure}[t]
%\vspace{-2mm}
\centering
\centerline{\includegraphics[width=\linewidth]{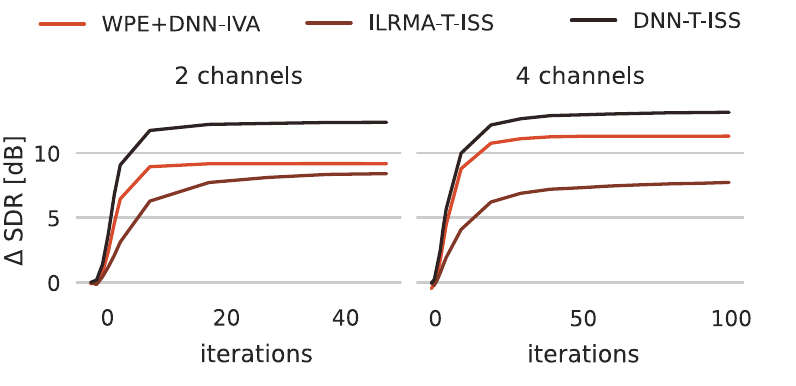}}
\vspace{-4mm}
\caption{Convergence of SDR improvement as a function of the number of the iterations on WSJ1-mix test set.}
\label{fig:sdr_convergence}
\vspace{-3mm}
\end{figure}
%------------------------------

Table~\ref{table:result} shows the evaluation results on the WSJ1-mix test set. 
The evaluation metrics were SDR, the short-time objective intelligibility (STOI)~\cite{stoi}, the perceptual evaluation of speech quality (PESQ)~\cite{pesq}, the speech-to-reverberation modulation energy ratio (SRMR)~\cite{srmr}, and WER.
Since DNN-MVDR estimateed two signals, it was evaluated with only two-speaker mixtures.
For two-speaker mixtures, our proposed method, DNN-T-ISS (marked with a $\star$), outperformed all the baselines on all the metrics.
The significant performance gap between ILRMA-T-ISS and DNN-T-ISS shows the effectiveness of learning the source model.
Also, our proposed method greatly outperformed the cascade approaches, which confirms the effectiveness of joint optimization of dereverberation and separation.
For three and four-speaker mixtures, our proposed method also gave the best performance.
In particular, our proposed method led to a significant improvement in WER, over 20\%, 30\% and 40\% reduction for two, three and four-speakers mixtures compared to the next best method, respectively.
This confirms that our proposed method can straightforwardly be applied to mixtures with different number of sources than during training.

Fig.\ref{fig:sdr_convergence} shows the convergence of SDR improvement as a function of the number of iterations on two and four-speaker mixtures.
Note that iterations of WPE is not included in Fig.\ref{fig:sdr_convergence}.
For both case, the neural source model led to faster convergence than the conventional source model.
In particular, in two-speaker case, our proposed method converged in 20 iterations, whereas ILRMA-T-ISS took approximately 40 iterations.

\subsection{Results on LibriCSS dataset}
\label{sec:results_libri}

%------------------------------ 
\begin{table}[t]
\begin{center}
\caption{WERs evaluated with LibriCSS dataset. Training is done with two-speaker mixtures from WSJ1-mix dataet. 0S/0L are 0\% overlap case with short/long inter-utterance silence.}
\label{table:libricss_result}
\vspace{-3.5mm}
\scriptsize
\scalebox{0.9}{
\begin{tabular}{@{}llr@{}}
\toprule
    {Ch.} & {Algo.} &{\begin{tabular}{@{}rrrrrrr@{}} \multicolumn{7}{c}{WER per overlap ratio in \%}\\{0S\;\;} & {0L\,} &{10.0} & {20.0} & {30.0} &{40.0} &{Avg}\end{tabular}}  \\

\midrule
  &Unprocess.~\cite{libricss} &\begin{tabular}{@{}rrrrrrr@{}}11.8  &11.7  &18.8  &27.2  &35.6  &43.3 &26.4\end{tabular} \\
\midrule
7 %&Chen \textit{et al.}~\cite{libricss} &\begin{tabular}{@{}rrrrrrr@{}}8.4\;\;  &8.3 &11.6 &15.8 &18.7 &21.7 &14.8 \end{tabular} \\
  &Chen \textit{et al.}~\cite{libricss_conformer} &\begin{tabular}{@{}rrrrrrr@{}}7.2\;\;  &7.5\;\;  &9.6 &11.3 &13.7 &15.1 &11.2 \end{tabular} \\
  &Wang \textit{et al.}~\cite{libricss_sota} &\begin{tabular}{@{}rrrrrrr@{}}5.8\;\;  &5.8\;\;  &5.9\,\,\, &6.5\;\, &7.7\;\, &8.3\;\,\, &6.8 \end{tabular} \\
\midrule

2 &WPE+DNN-MVDR      &\begin{tabular}{@{}rrrrrrr@{}}8.8\;\;  &8.9  &12.9 &18.6 &23.6 &29.0 &18.0\end{tabular}  \\
  &WPE+DNN-IVA       &\begin{tabular}{@{}rrrrrrr@{}}7.9\;\;  &8.1  &12.3 &17.2 &21.8 &27.1 &16.7\end{tabular}  \\
  &ILRMA-T-ISS       &\begin{tabular}{@{}rrrrrrr@{}}7.5\;\;  &7.7  &12.2 &17.1 &22.2 &26.5 &16.6\end{tabular} \\
  &DNN-T-ISS$^\star$ &\begin{tabular}{@{}rrrrrrr@{}}\bf{7.3}\;\;  &\bf{7.5}  &\bf{10.6} &\bf{14.3} &\bf{18.0} &\bf{21.2} &\bf{13.9}\end{tabular}\\
\hline

3 &WPE+DNN-MVDR      &\begin{tabular}{@{}rrrrrrr@{}}10.0     &10.0 &11.9 &14.8 &18.6 &21.4 &15.1\end{tabular}  \\
  &WPE+DNN-IVA       &\begin{tabular}{@{}rrrrrrr@{}}8.0\;\;  &8.0\;\;  &9.9  &13.0 &15.8 &18.0 &12.7\end{tabular} \\
  &ILRMA-T-ISS       &\begin{tabular}{@{}rrrrrrr@{}}6.6\;\;  &6.6\;\;  &9.2  &13.3 &17.0 &20.2 &12.9\end{tabular} \\
  &DNN-T-ISS$^\star$ &\begin{tabular}{@{}rrrrrrr@{}}\bf{6.5}\;\;  &\bf{6.3}\;\;  &\bf{7.6}  &\bf{10.8} &\bf{13.5} &\bf{13.9} &\bf{10.2}\end{tabular} \\
\hline

7 &WPE+DNN-MVDR      &\begin{tabular}{@{}rrrrrrr@{}}11.6 &12.5 &13.6 &16.4 &19.5 &21.0 &16.2\end{tabular}  \\
  &WPE+DNN-IVA       &\begin{tabular}{@{}rrrrrrr@{}}8.7\;\;  &8.9  &10.5 &12.8 &15.9 &17.4 &12.8\end{tabular} \\
  &ILRMA-T-ISS       &\begin{tabular}{@{}rrrrrrr@{}}\bf{6.6}\;\;  &6.9\;\;  &\bf{8.5} &12.9 &15.4 &18.1 &12.0\end{tabular} \\
  &DNN-T-ISS$^\star$ &\begin{tabular}{@{}rrrrrrr@{}}6.9\;\;  &\bf{6.6}\;\;  &8.9 &\bf{10.9} &\bf{13.6} &\bf{14.1} &\bf{10.6}\end{tabular} \\

\bottomrule
%\end{tabularx}
\end{tabular}
}
\end{center}
\vspace{-6mm}
\end{table}

% memo
% - params of GLU source model: 2.18 M
% - params of Conformer: 58.72 M
% - params of Wang et al. : 13.8 M ??

%%------------------------------

Table~\ref{table:libricss_result} shows the evaluation results on the LibriCSS.
Focusing on the separation with two channels, i.e., the same condition as training, our proposed method achieved the best performance.
For all overlap ratios, WPE+DNN-MVDR performed poorly compared to other baselines.
This may be due to the configuration of the DNN-MVDR such that it outputs two masks at the same time and \textit{over-separates} the sources.
In contrast, our proposed method, which learns a source prior for a \textit{single} source, worked well, although training set consisted of fully-overlapped mixtures, not containing overlap-free data.
This result gave a new insight that our proposed method can also be straightforwardly applied to low-overlap mixtures including overlap-free cases.
Compared to two channels case, using more channels, i.e., over-determined separation, significantly boosted the performance.
Using three channels led to as much as 3.7\% reduction of WER in our proposed method.
Although using all seven channels did not improve the performance over three channels case due to a slight over-separation, it still achieved much better performance than when using only two channels.

We also compare our proposed method with several prior works, Conformer-based MVDR beamforming with 58.72M parameters~\cite{libricss_conformer}, and deep convolutional network-based MVDR beamforming followed by another enhancement network with 13.8M parameters~\cite{libricss_sota}.
Note that the training set in~\cite{libricss_sota} were simulated assuming the knowledge of the target domain, e.g., the microphone position was the same as LibriCSS and the average overlap ratio was relatively low, about 33\%.
Compared to the Conformer-based approach~\cite{libricss_conformer}, which was trained with more than twice as much data as ours, the proposed method achieved better performance for all overlap ratios with much smaller number of parameters.
Although the proposed method did not reach the state-of-the-art~\cite{libricss_sota}, it showed promising performance without any knowledge of the target domain and with fewer number of parameters and microphones.

\subsection{Effectiveness of DMC}
\begin{figure}[t]
\begin{minipage}[b]{0.6\linewidth}
  \centering
  \centerline{\includegraphics[width=4.1cm]{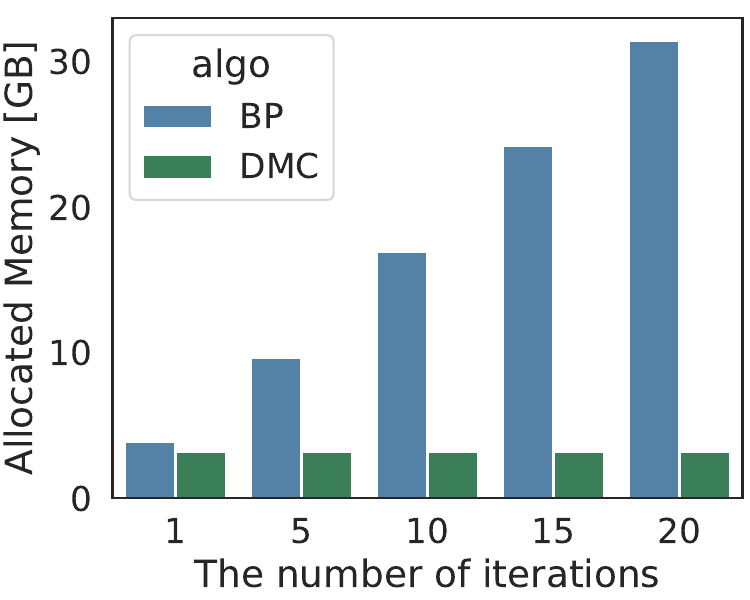}}
  \centerline{(a) Allocated memory}\medskip
\end{minipage}
\hfill
\begin{minipage}[b]{0.3\linewidth}
  \centering
  \centerline{\includegraphics[width=2.5cm]{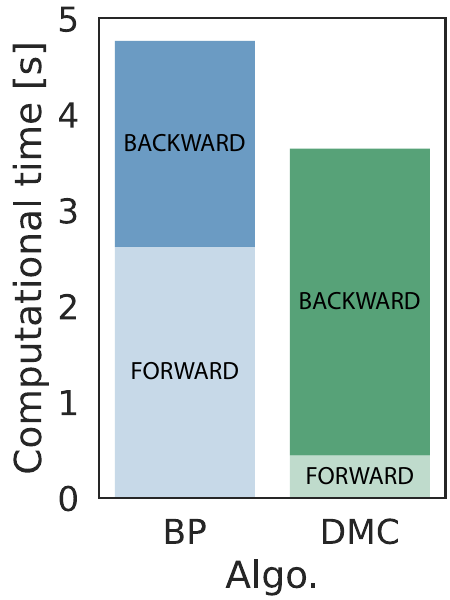}}
  \centerline{(b) Comp. time}\medskip
\end{minipage}
\vspace{-4mm}
\caption{(a) Allocated memory of BP and DMC. (b) Computational time for forward and backward pass, where batch size is 8 and the number of iterations is 20.}
\label{fig:memory_and_speed}
\vspace{-2mm}
\end{figure}

Fig.~\ref{fig:memory_and_speed} shows the memory cost as a function of the number of iterations (left), and runtimes of the forward and backward passes per batch of 8 samples (right).
Under the setting of our experiment, DMC reduced the memory cost from \SI{31}{\giga\byte} to only \SI{3}{\giga\byte}.
In addition, surprisingly, even though the backward of DMC is obviously slower, DMC was found to be faster than BP in the total of the forward and the backward passes.
Although we set the training parameters that allowed for learning without DMC, there are several advantages using DMC.
To begin with, it enables training with limited resources without decreasing the number of iterations or input signal length.
In preliminary experiments, we found out that reducing them makes training unstable.
In addition, the batch size can be enlarged, which could also contribute to stable training.
We investigate the performance with larger batch size in future work.

% -------------------------------------------------------------------------
% -------------------------------------------------------------------------

\section{Conclusion}
\label{sec:conclusion}
We proposed a independence-based joint dereverberation and separation method with a neural source model.
The source model was trained in an end-to-end manner directly with permutation invariant loss at the final output in the framework of T-ISS.
We also introduced DMC to reduce training memory cost of iterative approach.
We analyzed our proposed method from a variety of perspectives using in-domain fully overlapped synthetic mixtures and out-of-domain low-overlap real-world recordings.
In experiments, we showed that our proposed method worked better than the conventional model and the cascade connection of dereverberation and separation.
High performance was achieved regardless of the number of speakers or channels, including one-speaker case.
Furthermore, we demonstrated that our proposed method trained with fully overlapped synthetic mixtures also performed well on low-overlap real-world recordings without any modification.

\pagebreak

\bibliographystyle{IEEEtran}

\bibliography{template}

\end{document}